\documentclass{optica-article}

\journal{opticajournal} 

\articletype{Research Article}

\usepackage{lineno}

\usepackage{caption}
\usepackage{subcaption}
\usepackage{comment}
\usepackage{booktabs}

\begin{document}

\title{A two-dimensional gallium phosphide optomechanical crystal in the resolved-sideband regime}

\author{Sho Tamaki\authormark{1,2}, Mads Bjerregaard Kristensen\authormark{1,2}, Théo Martel\authormark{3}, Rémy Braive\authormark{3,4,5}, and Albert Schliesser\authormark{1,2,*}}

\address{\authormark{1}Niels Bohr Institute, University of Copenhagen, Blegdamsvej 17, Copenhagen 2100, Denmark\\
\authormark{2}Center for Hybrid Quantum Networks, Niels Bohr Institute, University of Copenhagen, Blegdamsvej 17, Copenhagen 2100, Denmark\\
\authormark{3}Centre de Nanosciences et de Nanotechnologies, CNRS, Université Paris Saclay, Palaiseau 91120, France\\
\authormark{4}Université Paris Cité, Paris 75006, France, France\\
\authormark{5}Institut Universitaire de France (IUF), Paris 75231, France}

\email{\authormark{*}albert.schliesser@nbi.ku.dk} 


\begin{abstract*} 
Faithful quantum state transfer between telecom photons and microwave frequency mechanical oscillations necessitate a fast conversion rate and low thermal noise.
Two-dimensional (2D) optomechanical crystals (OMCs) are favorable candidates that satisfy those requirements.
2D OMCs enable sufficiently high mechanical frequency (1$\sim$10 GHz) to make the resolved-sideband regime achievable, a prerequisite for many quantum protocols.
It also supports higher thermal conductance than 1D structures, mitigating the parasitic laser absorption heating.
Furthermore, gallium phosphide (GaP) is a promising material choice thanks to its large electronic bandgap of 2.26 eV, which suppresses two-photon absorption, and high refractive index $n$ = 3.05 at the telecom C-band, leading to a high-$Q$ optical mode.
Here, we fabricate and characterize a 2D OMC made of GaP.
We realize a high optical $Q$-factor of $7.9\times 10^{4}$, corresponding to a linewidth $\kappa/2\pi$ = 2.5 GHz at the telecom frequency 195.6 THz.
This optical mode couples to several mechanical modes, whose frequencies all exceed the cavity linewidth.
The most strongly coupled mode oscillates at 7.7 GHz, more than 3 times the optical linewidth, while achieving a substantial vacuum optomechanical coupling rate $g_{\mathrm{0}}/2\pi$ = 450 kHz.
This makes the platform a promising candidate for a long-lived, deterministic quantum memory for telecom photons at low temperatures.

\end{abstract*}

\section{Introduction}
\begin{figure}[ht!]
\centering
\includegraphics[width=\linewidth]{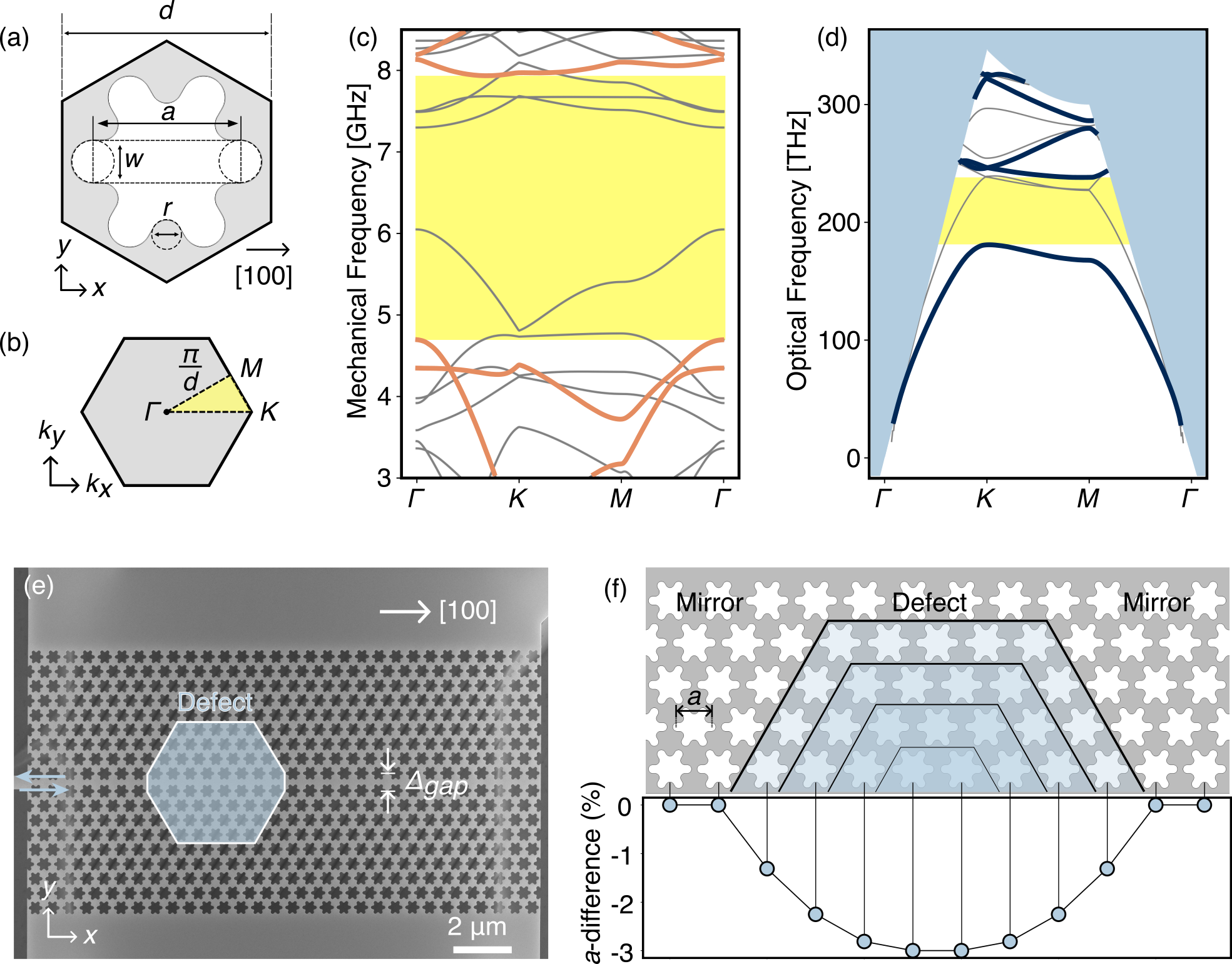}
\captionsetup{width=1\linewidth}
\caption{
Description of the OMC design. (a) Geometry of a unit cell. $( a,w,r,d )$ = (370, 120, 80, 577) nm and the thickness of 260 nm. The direction of the [100] crystalline axis is indicated by an arrow. (b) The 1st Brillouin zone. Band diagrams are simulated along the line surrounding the area indicated in yellow in the order of $\Gamma\rightarrow K \rightarrow M \rightarrow \Gamma$. 
(c) Mechanical band diagram. Relevant modes ($yz$-symmetric at the $\Gamma$ point) are highlighted by orange lines. The bandgap (4.77-7.94 GHz) is indicated in yellow.
(d) Optical band diagram. TE modes are highlighted by blue lines. The bandgap (181-238 THz) is indicated in yellow. The blue-shaded area shows the light cone. (e) SEM image of the full structure. The [100] crystalline axis lies parallel to the $x$-axis. The defect region is highlighted in blue. The horizontal gap, $\Delta_{\mathrm{gap}}$=600 nm, in the middle of the structure enables waveguide-like modes. (f) Schematic of the defect design. The $a$-parameter of snowflakes varies quadratically with a 3\% difference from the bulk value.}
\label{fig:design}
\end{figure}
Nanomechanical systems in the quantum regime have inspired researchers and engineers due to their potential applications ranging from ultra-sensitive force sensing to quantum communication\cite{gisler2024enhancingmembranebasedscanningforce,meesala_non-classical_2024}.
One specific application is an optomechanical quantum memory which uses a mechanical mode as a quantum coherent memory for telecom photons\cite{PhysRevLett.132.100802,Andreas2020,Fiaschi2021}, motivated by the long coherence time of micro-/nano-mechanical vibrations interacting effectively with telecom photons.
Quantum memory is a building block of the quantum repeater protocol\cite{Briegel-repeater-1998} that can surpass the current limit on transportation distance of optical quantum information through optical fibers ($\sim$600 km\cite{pittaluga_600-km_2021}).
The first quantum memory and repeater for light have been experimentally demonstrated with atomic gasses\cite{julsgaard_experimental_2004,yuan_experimental_2008}, and recently, optomechanical systems have achieved storage for telecomband wavelengths\cite{Andreas2020,Fiaschi2021}.
However, deterministic write/readout of optical states has yet to be demonstrated for practical quantum memories.

Deterministic storage requires a fast optomechanical conversion rate that matches the bandwidth of incoming photons ($\sim$200 MHz\cite{pedersen_near_2020}).
Generally speaking, the conversion rate scales with the number of intra-cavity photons and thus the intensity of the input laser\cite{aspelmeyer_cavity_2014}.
However, laser-absorption-induced heating at millikelvin temperature sets an upper limit for input laser power as it significantly degrades mechanical coherence.
This results in a trade-off between the conversion efficiency and the amount of thermal noise, thus restricting the performance of mechanical quantum memories and transducers\cite{Andreas2020,jiang_optically_2023,meesala_non-classical_2024}.
Accordingly, an optomechanical system robust against laser absorption heating is strongly desired.

Large efforts have been made in structural design to mitigate the problem.
It has been discovered that 2D structures can significantly reduce the laser-absorption-induced thermal bath occupancy compared to 1-dimensional structures\cite{HRen2020}.
Also, a sophisticated waveguide design that detaches the 2D mechanical mode from the local heat bath is capable of enhancing the performance of optomechanical transducers\cite{sonar_high-efficiency_2024}.
Another direction of effort is to seek a suitable material that inherently absorbs less laser light.
Recent studies \cite{Schneider2019, Stockill2019} have suggested gallium phosphide (GaP) as a favorable candidate due to (i) the large electronic bandgap of 2.26 eV exceeding the energy of two telecom photons ($\sim$1.6 eV), which can suppress the two-photon-absorption-induced thermalization, and (ii) the high refractive index $n=3.05$ enabling a high-$Q$ optical mode and large vacuum optomechanical coupling $g_{0}$.

Here, we design and characterize a 2D GaP optomechanical crystal (OMC).
Our device hosts an optical resonance at 
$\omega_{\mathrm{o}}/2\pi$ = 195.55 THz, $\lambda_{\mathrm{o}}$ = 1533.1 nm with linewidth $\kappa/2\pi$ = 2.47 GHz which couples to 3 localized mechanical modes.
The most strongly coupled mechanical mode has a frequency $\Omega_{\mathrm{m}}/2\pi=$ 7.7 GHz and intrinsic linewidth $\Gamma_{\mathrm{m}}/2\pi$ = 4.9 MHz at room temperature, placing our device well in the resolved-sideband regime ($\Omega_{\mathrm{m}}>3\kappa$).
From a thermally calibrated measurement\cite{Gorodetsky2010} with different detunings, we infer a vacuum optomechanical coupling rate $g_0/2\pi$ = 452 kHz for this mode.

\section{Methods}
\begin{figure}[t]
\centering
\includegraphics[width=\linewidth]
{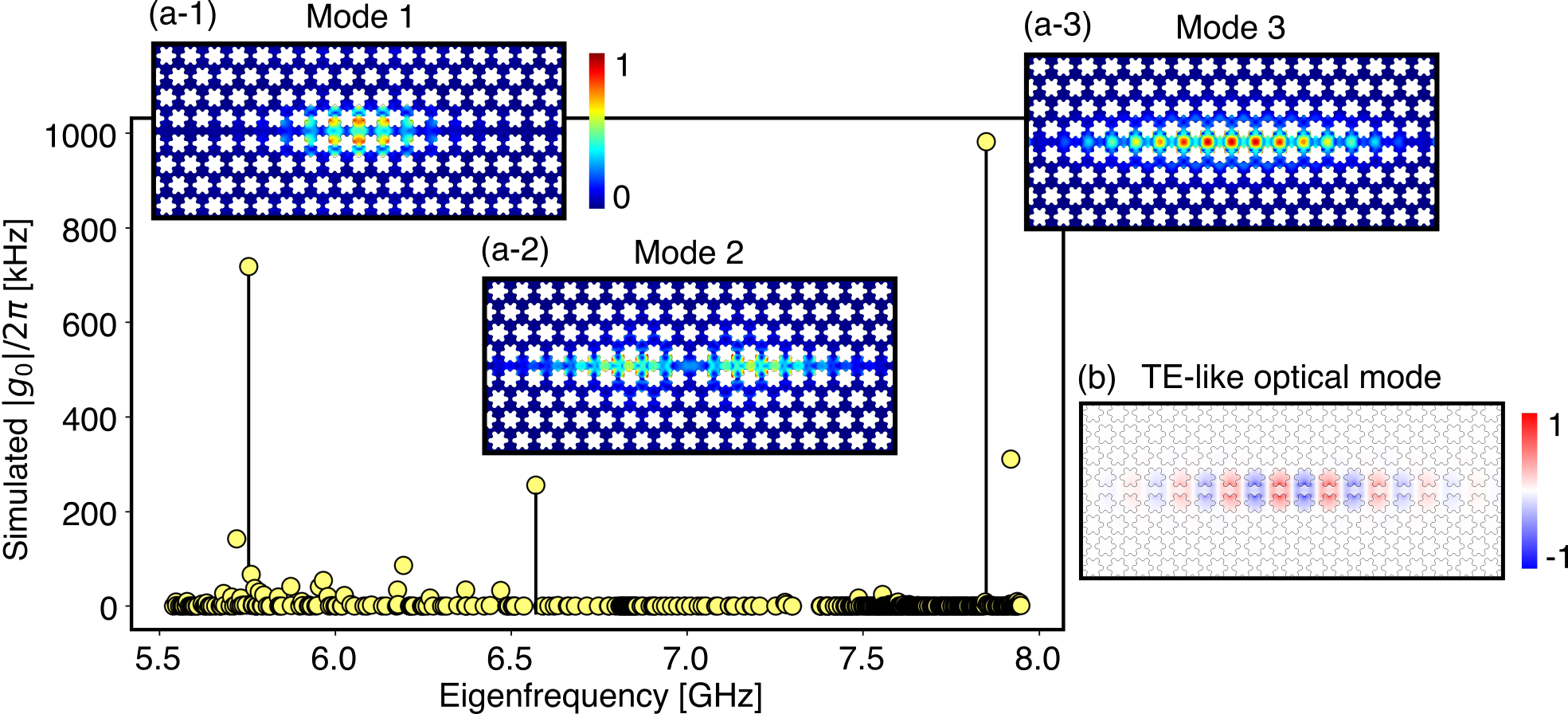}
\captionsetup{width=1\linewidth}
\caption{
Simulated $|g_{0}|$ for each mechanical eigenmode. Three dominant modes with high coupling rates are indicated by black lines at (5.75, 6.57, 7.85) GHz.
The inset (a) shows the absolute mechanical displacement of each mode.
(b) $E_{y}$ of the TE-like optical cavity mode with simulated frequency $\omega_{\mathrm{o}}/2\pi=192.7$ THz, which couples to those mechanical modes.
}
\label{fig:full-simu}
\end{figure}
Our OMC has snowflake-shaped holes\cite{Safavi-Naeini2014,PhysRevApplied.21.014015} in a periodic hexagonal lattice over the suspended GaP slab that opens bandgaps for both mechanical and optical modes. Fig$.$ \ref{fig:design}(a)
shows the dimension of a unit cell with parameters of $(a,w,r,d)$ = (370, 120, 80, 577) nm and a thickness of 260 nm.
This shape has more degrees of freedom for optical and mechanical mode engineering than the typical ellipses.
As for the mechanical properties, GaP has a Zincblende crystalline structure  $\bar{4}3m$, giving rise to anisotropy of mechanical properties with 4-fold rotational symmetry.
Therefore, the mechanical modes depend on the alignment of the OMC structure to the crystalline axis.
We align the snowflake so that one of the spikes points to the [100] crystalline axis to maximize the mechanical frequency while maintaining sufficient coupling strength.
We use elastic tensor components of $(C_{11},C_{12},C_{44})$ = (140.5, 62.0, 70.3) GPa for our simulation.

The band diagrams for both modes are simulated over $\Gamma\rightarrow K \rightarrow M \rightarrow \Gamma$ in the 1st Brillouin zone which is shown in Fig$.$ \ref{fig:design}(b,c).
As for the mechanical modes, we focus only on the modes with $yz$-symmetry at $\Gamma$ point, sharing parity with the optical mode.
Any mode without this symmetry results in small optomechanical coupling as the contribution from one side is canceled from the opposite side of the device.
We find our design has a wide bandgap for relevant modes > 3 GHz around $\sim$6.5 GHz as well as a full bandgap > 1 GHz.
At the same time, the optical bands have a gap between around the telecom C-band (181-238 THz).

We then design the full device structure based on the unit cell simulations.
Subtracting a row of snowflakes, and adding a gap $\Delta_{\mathrm{gap}}$ = 600 nm shown in Fig$.$ \ref{fig:design}(e) creates waveguide modes with one-dimensional confinement.
The additional quadratic variation in snowflake size ($a$-parameter of the geometry) forms the defect region leading to two-dimensional confinement.
The gradual parameter variation gently shifts the bandgap from the mirror to the defect region resulting in high-$Q$ modes\cite{akahane_high-q_2003}.
Our device has a defect region formed over 8 snowflakes with the minimum $a$-parameter being only 3\% smaller than that of the mirror region as presented in Fig$.$ \ref{fig:design}(f).

We simulate optical and mechanical modes numerically with the finite element solver of COMSOL Multiphysics\cite{comsol} as shown in Fig$.$ \ref{fig:full-simu}.
The simulated TE-like optical mode Fig$.$ \ref{fig:full-simu}(b) has a telecom C-band frequency $\omega_{\mathrm{o}}/2\pi=192.7$ THz.
The vacuum optomechanical coupling rate $g_{0}$ is then computed by spatially integrating both modes considering the moving boundary (MB) and photoelastic (PE) contributions\cite{chan_optimized_2012,balram_moving_2014}.
MB contribution comes from the surface integral of the product of the electric field and the surface displacement, while PE is calculated by the volume integral of the product of the electric field and strain.
Simulated $g_{0}$'s are plotted in Fig$.$ \ref{fig:full-simu}(c) for each mechanical eigenmode.
There are 3 dominant mechanical modes at (5.75, 6.57, 7.85) GHz with sizable couplings $g_{0}/2\pi$ = (-718, -256, -983) kHz, with MB and PE contributions of $g_{\mathrm{mb}}/2\pi$ = (-305, -156, -755) kHz and $g_{\mathrm{pe}}/2\pi$ = (-414, -101, -228) kHz, respectively.
Note that the sign of the total coupling rate does not affect the characterization of the optomechanical device in this work.
We find both MB and PE contribute comparably to these modes thanks to the relatively high photoelastic constants of GaP $(p_{11}, p_{12}, p_{44})$ = (-0.23, -0.13, -0.10)\cite{mytsyk_elasto-optic_2015}, which are a factor of 2 larger than those of silicon\cite{biegelsen_photoelastic_1974}.

\begin{figure}[t]
\centering
\includegraphics[width=\linewidth]{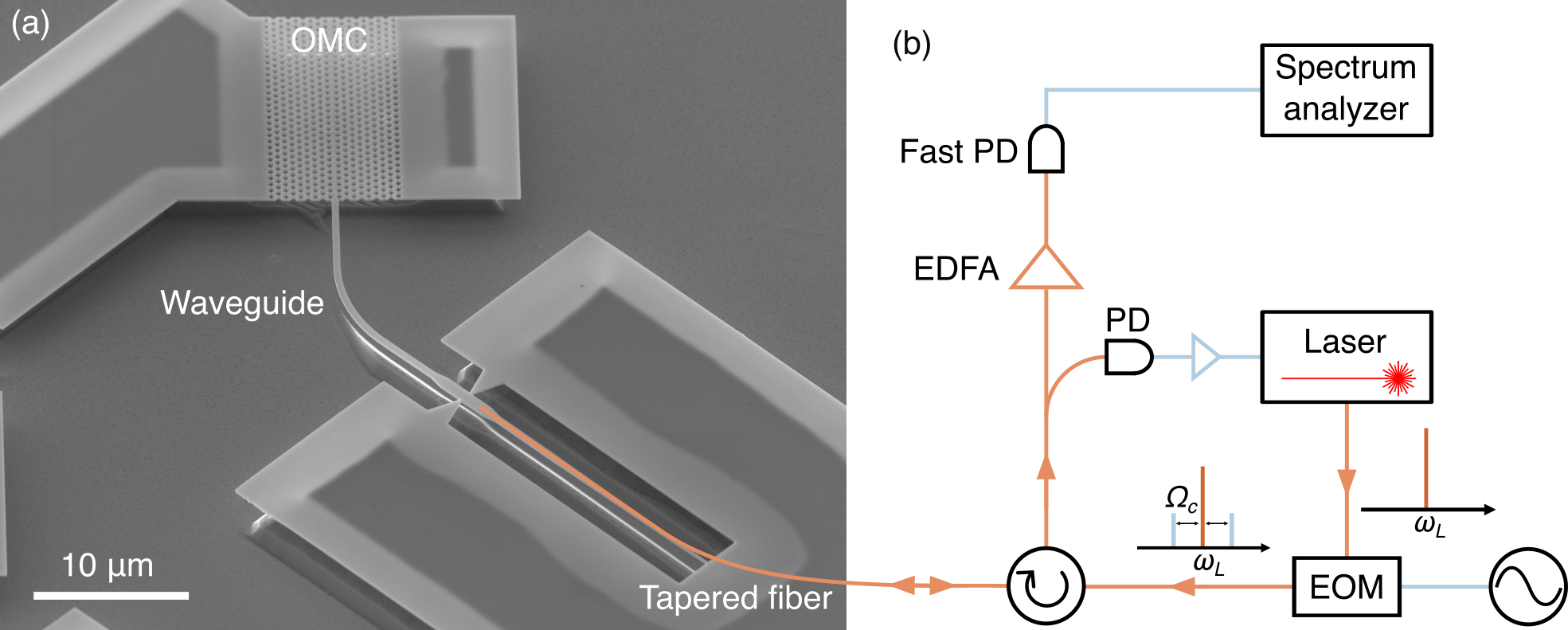}
\captionsetup{width=1\linewidth}
\caption{
Experimental setup. (a) SEM image of the OMC and waveguide for evanescent fiber coupling. The orange line indicates the fiber position. (b) Schematic of our measurement setup for device characterization. Orange and blue lines denote optical and electrical paths respectively. EOM: electro-optic phase modulator, EDFA: erbium-doped fiber amplifier, PD: photodetector, $\omega_{\mathrm{L}}$: laser frequency, and $\Omega_{\mathrm{c}}$: modulation frequency.
}
\label{fig:setup}
\end{figure}
The devices are fabricated on a 260-nm-thick GaP layer epitaxially grown on a gallium arsenide substrate (see supplemental for further details).
Fig$.$ \ref{fig:setup}(a) shows a scanning electron micrograph of the entire OMC and the access waveguide.
The 500 nm wide rectangular waveguide supports a single TE traveling mode with an effective index of 1.99.
We employ evanescent coupling between the waveguide and a tapered fiber\cite{tiecke_efficient_2015,burek_fiber-coupled_2017} for optical access (see supplemental for further details).
This configuration will allow us to make cryogenic packaging of the device\cite{zeng_cryogenic_2023} while suppressing the vibrational noise from the coupling fiber.
Our setup achieves a coupling efficiency as high as > 60\%.
Devices are characterized with the measurement setup shown in Fig$.$ \ref{fig:setup}(b).
The input light field comes from a tunable external cavity diode laser at frequency $\omega_{\mathrm{L}}$.
An electro-optic phase modulator (EOM) creates modulation sidebands at $\omega_{\mathrm{L}}\pm\Omega_{\mathrm{c}}$ used to determine the vacuum optomechanical coupling rate $g_{0}$.
A portion of the reflected light is sent to a photodetector (PD) to measure the optical spectrum and lock the laser detuning $\Delta$.
The other part of the reflected light is amplified by an erbium-doped fiber amplifier (EDFA) and then detected by a fast PD whose microwave signal is sent to a spectrum analyzer to characterize mechanical spectral properties.
All characterization is carried out at room temperature and ambient pressure.

\section{Results}

\begin{figure}[t]
\centering
\includegraphics[width=\linewidth]{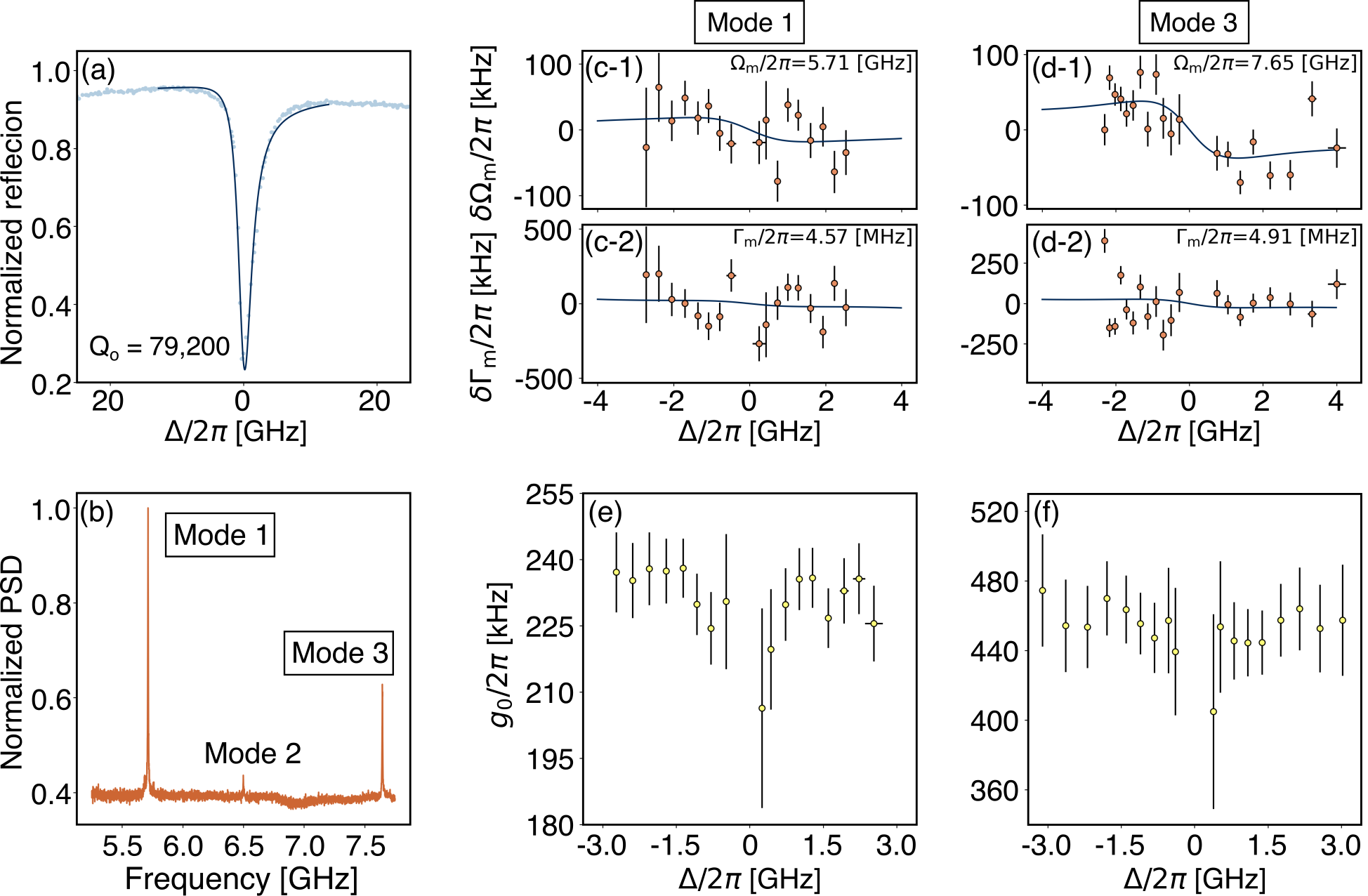}
\captionsetup{width=1\linewidth}
\caption{
Characterization results. (a) Optical spectrum around resonance frequency $\omega_{\mathrm{o}}/2\pi$ = 195.55 THz, wavelength $\lambda_{\mathrm{o}}$ = 1533.1 nm showing a loaded quality factor $Q_{\mathrm{o}}$ = $7.92\times10^4$ ($\kappa/2\pi$ = 2.47 GHz). (b) Mechanical power spectral density (PSD) demonstrating 3 dominant modes around frequencies expected from the simulation. (c,d) Measured mechanical frequency and linewidth shift over detunings showing the dynamical backaction effect for modes 1 and 3. Estimated intrinsic frequency and linewidth are ($\Omega_{\mathrm{m}}/2\pi$, $\Gamma_{\mathrm{m}}/2\pi$) = (5.71 GHz, 4.57 MHz) for mode 1 and (7.65 GHz, 4.91 MHz) for mode 3. (e,f) Vacuum optomechanical coupling rate for each mode, yielding $g_{0}/2\pi$ = 231 (452) kHz for mode 1 (3), respectively.
}
\label{fig:result1}
\end{figure}
The optical spectrum is measured by scanning the laser frequency while acquiring the reflected optical power with a PD.
The laser power is set to a sufficiently small level to suppress the thermo-optical nonlinearity\cite{carmon_dynamical_2004}, which shifts the optical resonance towards a lower frequency depending on the power of a laser.
Fig$.$ \ref{fig:result1}(a) shows the measured optical reflection and the fitting curve
\begin{align}\label{eq:fitfunc}
    R(\Delta) = h - A\frac{(1-q^2)\kappa/2-q\Delta}{\kappa^2/4+\Delta^2},
\end{align}
derived from a model accounting for spurious back-reflection from the waveguide, detailed in the supplemental material.
Here, we define the offset $h$, amplitude parameter $A$, and the Fano parameter $q$, in addition to linewidth $\kappa$ and detuning $\Delta$.
For optical resonance at $\lambda_{\mathrm{o}}$ = 1533.1 nm,  $\omega_{\mathrm{o}}/2\pi$ = 195.55 THz, this gives the loaded quality factor $Q_{\mathrm{o}}$ = $7.92\times10^4$ (linewidth $\kappa/2\pi$ = 2.47 GHz).
The output field is intensity-modulated at the mechanical frequency $\Omega_{\mathrm{m}}$ due to the combination of phase modulation by mechanical displacement and quadrature rotation by the optical cavity.
Fig$.$ \ref{fig:result1}(b) plots the power spectral density obtained by the spectrum analyzer.
We find 3 mechanical peaks at around $\Omega_{\mathrm{m}}/2\pi$ = 5.7, 6.5, 7.7 GHz, as predicted from the simulation.

To determine the precise values of the mechanical parameters, we analyze the dynamical backaction effects\cite{aspelmeyer_cavity_2014} on modes 1 and 3 by changing laser detuning $\Delta$ at fixed input power.
Measured shifts on mechanical frequencies and linewidths are plotted and fitted in Fig$.$ \ref{fig:result1}(c,d). The fitting curves are
\begin{align}
    \Omega_{\mathrm{eff}} &= \Omega_{\mathrm{m}} + \delta \Omega_{\mathrm{m}},\\
    \delta \Omega_{\mathrm{m}} &= g_{0}^{2}\frac{\kappa_{\mathrm{ex}}}{\Delta^{2}+(\kappa / 2)^{2}} \frac{P}{\hbar \omega_{\mathrm{L}}} \left(\frac{\Delta-\Omega_{\mathrm{m}}}{\kappa^{2} / 4+\left(\Delta-\Omega_{\mathrm{m}}\right)^{2}}+\frac{\Delta+\Omega_{\mathrm{m}}}{\kappa^{2} / 4+\left(\Delta+\Omega_{\mathrm{m}}\right)^{2}}\right),\\
    \Gamma_{\mathrm{eff}} &= \Gamma_{\mathrm{m}} + \delta \Gamma_{\mathrm{m}},\\
    \delta \Gamma_{\mathrm{m}}&=  g_{0}^{2}\frac{\kappa_{\mathrm{ex}}}{\Delta^{2}+(\kappa / 2)^{2}} \frac{P}{\hbar \omega_{\mathrm{L}}}\left(\frac{\kappa}{\kappa^{2} / 4+\left(\Delta+\Omega_{m}\right)^{2}}-\frac{\kappa}{\kappa^{2} / 4+\left(\Delta-\Omega_{m}\right)^{2}}\right),
\end{align}
where we define intrinsic frequency $\Omega_{\mathrm{m}}$, linewidth $\Gamma_{\mathrm{m}}$, input power to the optical cavity $P$, and external cavity loss $\kappa_{\mathrm{ex}}$.
We extract ($\Omega_{\mathrm{m}}/2\pi,\Gamma_{\mathrm{m}}/2\pi$) = ( 5.71 GHz, 4.57 MHz) for mode 1 and (7.65 GHz, 4.91 MHz) for mode 3.
Both modes are sideband-resolved ($\kappa < \Omega_{\mathrm{m}}$) which is a prerequisite for an effective optomechanical memory.
Note that the linewidth shifts for both modes are somewhat unclear because they are well sideband-resolved and the shifts are quite small in the region of $|\Delta|<\Omega_{\mathrm{m}}$.

The optomechanical coupling rate $g_{0}$ is estimated by comparing the area below the mechanical noise power spectral density and the delta-function-like calibration tone generated by the EOM, whereby $g_{0}$ is calculated as\cite{Gorodetsky2010}:
\begin{align}
    g_{0} \approx \sqrt{ \frac{\phi_{0}^{2} \Omega_{\mathrm{c}}^{2}}{4\left\langle n_{\mathrm{th}}\right\rangle} \frac{S_{I I}^{\mathrm{meas}}\left(\Omega_{\mathrm{m}}\right) \Gamma_{\mathrm{m}} / 4}{S_{I I}^{\mathrm{meas}}\left(\Omega_{\mathrm{c}}\right) f_\mathrm{ENBW}}},
\end{align}
where $\phi_0$ is the phase modulation depth, $f_{\mathrm{ENBW}}$ the effective noise bandwidth of the spectrum analyzer in linear frequency, $S_{II}^{\mathrm{meas}}(\Omega)$ the symmetrized noise power spectral density at $\Omega$, and thermal mechanical occupation $\left\langle n_{\mathrm{th}}\right\rangle \approx k_{\mathrm{B}}T/ \hbar \Omega_{\mathrm{m}}$.
The modulation frequency $\Omega_{\mathrm{c}}/2\pi$ is set to no further from $\Omega_{\mathrm{m}}/2\pi$ than 40 MHz.
We measure $g_0$ in different laser detunings plotted in Fig$.$ \ref{fig:result1}(e,f) and obtain $g_0 /2\pi$ = 231±13 (452±32) kHz for mode 1 (3).

\begin{table}[t]
\centering
\resizebox{\textwidth}{!}{\begin{tabular}{|l|cc|ccccc|}
\hline
\rowcolor[HTML]{b3cde0}
Mode& \multicolumn{1}{l}{Simulated}&  &\multicolumn{1}{l}{Measured}  & & & & \\ \hline
&$\Omega_{\mathrm{m}}/2\pi$ [GHz] & $|g_0|/2\pi$ [kHz]&   $\Omega_{\mathrm{m}}/2\pi$ [GHz]& $\Gamma_{\mathrm{m}}/2\pi$ [MHz]& $\Omega_{\mathrm{m}}/\kappa$ & $g_{0}/2\pi$ [kHz] & $C_\mathrm{0}=4g_{0}^2/\Gamma_{\mathrm{m}}\kappa$ \\ \hline
\multicolumn{1}{|l|}{This work}& & & & & & & \\
\multicolumn{1}{|c|}{Mode 1}& 5.75 & 719 &  5.71 $\pm$ 0.01 & 4.57 $\pm$ 0.04 & \textbf{2.31} & 231 $\pm$ 13&  $1.89\times10^{-5}$ \\
\multicolumn{1}{|c|}{Mode 2}& 6.57 & 257 &  6.50 $\pm$ 0.01 & 3.93 $\pm$ 0.41 & \textbf{2.63} & &\\
\multicolumn{1}{|c|}{Mode 3}& 7.85 & 983 &  7.65 $\pm$ 0.01 & 4.91 $\pm$ 0.03 & \textbf{3.10} & 452 $\pm$ 32 & $6.74\times10^{-5}$ \\ \hline
\multicolumn{1}{|l|}{Schneider {\it{et al.}}\cite{Schneider2019}}& 2.84 & 760 &  2.90 & 3.13 & 0.92 & 370-680 & $5.55$-$18.8\times10^{-5}$ \\ \hline
\multicolumn{1}{|l|}{Hönl {\it{et al.}}\cite{honl_microwave--optical_2022}}& & & & & &  & \\
\multicolumn{1}{|c|}{Mode A}& 3.17 & 173 &  3.28 & 2.55 & 1.12 & 191 & $1.96\times10^{-5}$ \\
\multicolumn{1}{|c|}{Mode B}& 3.23 & 381 &  3.31 & 2.81 & 1.13 & 283 & $3.90\times10^{-5}$ \\
\multicolumn{1}{|c|}{Mode C}& 3.25 & 506 &  3.33 & 2.56 & 1.14 & 293 & $4.59\times10^{-5}$ \\ \hline
\multicolumn{1}{|l|}{Stockill {\it{et al.}}\cite{Stockill2019}}& 2.85 & 525 &  2.91 & \textbf{0.014} (@ 7 mK) & 0.56 & \textbf{845} & $\boldsymbol{4.03\times10^{-2}}$ \\ \hline
\end{tabular}}
\captionsetup{width=1\linewidth}
\caption{Comparison of simulated and measured parameters from different GaP OMCs.}
\label{table:result}
\end{table}

Table \ref{table:result} compares parameters among GaP OMCs including this work and previously realized 1D devices.
Our mechanical modes show good agreement with simulations in their mechanical frequencies and they are 2 to 3 times higher than those of 1D OMCs of previous studies, thus realizing by far the largest $\Omega_{\mathrm{m}}/\kappa$ ratio of 3.10 for the mode 3.
The measured $g_0$'s are also comparably high.
We find that the measured $g_0$ for each mode is smaller than the simulated by a factor of $\sim$2-3.
Previous studies\cite{Schneider2019,honl_microwave--optical_2022} have also reported a factor of $\sim$2 difference.
We attribute this to our imperfect knowledge of the photoelastic constants of GaP in the telecom band. 
Additionally, the 2D structures' sensitivity to fabrication imperfection compared to 1D structures is still debatable (see supplemental for further details).
For the measured $g_{0}$ by Hönl {\it{et al.}}\cite{honl_microwave--optical_2022}, we take the mean value of blue- and red-detuned measurements.
We further compare the single-photon cooperativity $C_{\mathrm{0}}=4g_{0}^2/\Gamma_{\mathrm{m}}\kappa$, which corresponds to an optomechanical cooperativity per cavity photon.
Our device achieves $C_{\mathrm{0}}=6.74\times10^{-5}$ for mode 3 which is one of the highest among the room temperature GaP OMCs studied so far.
This is quite an encouraging result considering applications such as quantum memory.
As previously demonstrated in a silicon OMC\cite{HRen2020,sonar_high-efficiency_2024}, 2D structures have an advantage on thermalization at a given laser power.
Therefore, we may deduce that our device can achieve higher quantum cooperativity $C_{\mathrm{q}}=n_{\mathrm{cav}}C_{\mathrm{0}}/n_{\mathrm{th}}$ than 1D GaP OMC at cryogenic temperature as we can input stronger power, thus more cavity photons $n_{\mathrm{cav}}$.

\section{Conclusion}
We design and characterize a GaP optomechanical crystal with a 2D geometry.
The careful design realizes the highest $\Omega_{\mathrm{m}}/\kappa$ ratio > 3 yet achieved in GaP OMCs, which places well in the sideband-resolved regime.
This is one of the requirements for faithful photon-phonon conversion.
From dynamical backaction measurements, we precisely characterize the mechanical properties.
The most strongly coupled mode achieves a large vacuum optomechanical coupling rate $g_{0}/2\pi$=452 kHz to a telecom optical cavity mode.
The measured $g_{0}$'s are smaller compared to the simulation results by similar factors that have previously been reported in one-dimensional devices\cite{Schneider2019,honl_microwave--optical_2022}.
We speculate this is a common problem due to the imperfect knowledge of the photoelastic constant of GaP for telecom photons.
In addition, 2D designs may be more vulnerable to fabrication imperfections such as uneven sidewalls, because of the structural complexity.

We expect our device to be less susceptible to laser heating thanks to the 2D structural design and GaP material properties, compared to other OMC devices.
The design is expected to diffuse thermal phonons faster than 1D thus lowering absorption-induced thermal occupation as previously reported on silicon devices\cite{HRen2020}.
We further assume that the recently reported waveguide design\cite{sonar_high-efficiency_2024}, which successfully decouples the local waveguide heat bath from the OMC cavity, is capable of further enhancing the thermalization performance of our device.
Although GaP is a quite promising material considering its large electronic bandgap, additional thin Al$_2$O$_3$ layer deposition may result in better suppression of optical absorption at the surfaces as reported on GaAs OMCs\cite{forsch_microwave--optics_2020}.
On top of that, there is still room for improving the evanescent fiber-waveguide coupling efficiency, which is crucial for deterministic photon-phonon conversion, by optimizing the waveguide design and the fiber etching process.
A coupling efficiency of 97\% has been reported in a diamond nanobeam photonic crystal\cite{tiecke_efficient_2015}.
This coupling configuration will allow us to make a cryogenic packaging of the device for millikelvin temperature experiments\cite{zeng_cryogenic_2023}.
With those further improvements and the currently demonstrated optomechanical memory protocol\cite{PhysRevLett.132.100802}, our device opens the possibility for deterministic quantum memories for single telecom photons.

Previous studies on silicon OMCs\cite{HRen2020} have revealed the superior performance of 2D structures over 1Ds in terms of thermalization.
An absorption-induced thermal bath occupancy $n_{\mathrm{p}}=1.1\times n_{\mathrm{cav}}^{0.3}$ has been reported at cryogenic temperature ($\sim$10 mK) on a 2D device with the number of cavity photons of $n_{\mathrm{cav}}<10^4$.
This result is 7 times smaller than that of 1D OMC thanks to the enhanced thermal conductance.
Another study on GaP OMC\cite{Schneider2019} at room temperature has shown at least 5 times smaller optical absorption rate than silicon in the same photon number range.
Combining those naively suggests that our device will be allowed to host 5 times more cavity photons than 2D Silicon OMCs while suppressing the thermalization.
As the photon-phonon conversion rate scales with $g_{0}^{2}n_{\mathrm{cav}}/\kappa$, more photons are beneficial to increase the conversion rate and match the bandwidth of single-photons.
However, it should be noted that heating dynamics at cryogenic temperature is still not yet fully understood. The temperature rise at room temperature is linear to the input power\cite{Schneider2019} which is different from that is reported in cryogenic experiments\cite{HRen2020}.

\section*{Acknowledgements}
This work was supported by the European Research Council project PHOQS (Grant No. 101002179), the Novo Nordisk Foundation (Grant No. NNF20OC0061866), the Danish National Research Foundation (Center of Excellence “Hy-Q”), the Independent Research Fund Denmark (Grant No. 1026-00345B), the French RENATECH network, as well as the CNRS-KU grant (Grant No. C2N UMR9001).


\section*{Supplemental document}
See Supplement 1 for supporting content.


\newpage

\title{A two-dimensional gallium phosphide optomechanical crystal in the resolved-sideband regime: supplemental document}
\author{Sho Tamaki, Mads Bjerregaard Kristensen, Théo Martel, Rémy Braive, and Albert Schliesser} 

\begin{abstract*}
This document provides supplementary information to "A two-dimensional gallium phosphide optomechanical crystal in the resolved-sideband regime". It includes details on device fabrication and design tolerance to the fabrication imperfection, tapered fiber fabrication, polarization matching conditions affecting the determination of vacuum coupling rate $g_0$, and a theoretical approach to account for it.
\end{abstract*}



\section{Device fabrication}

\begin{figure}[ht!]
\centering
\includegraphics[width=\linewidth]{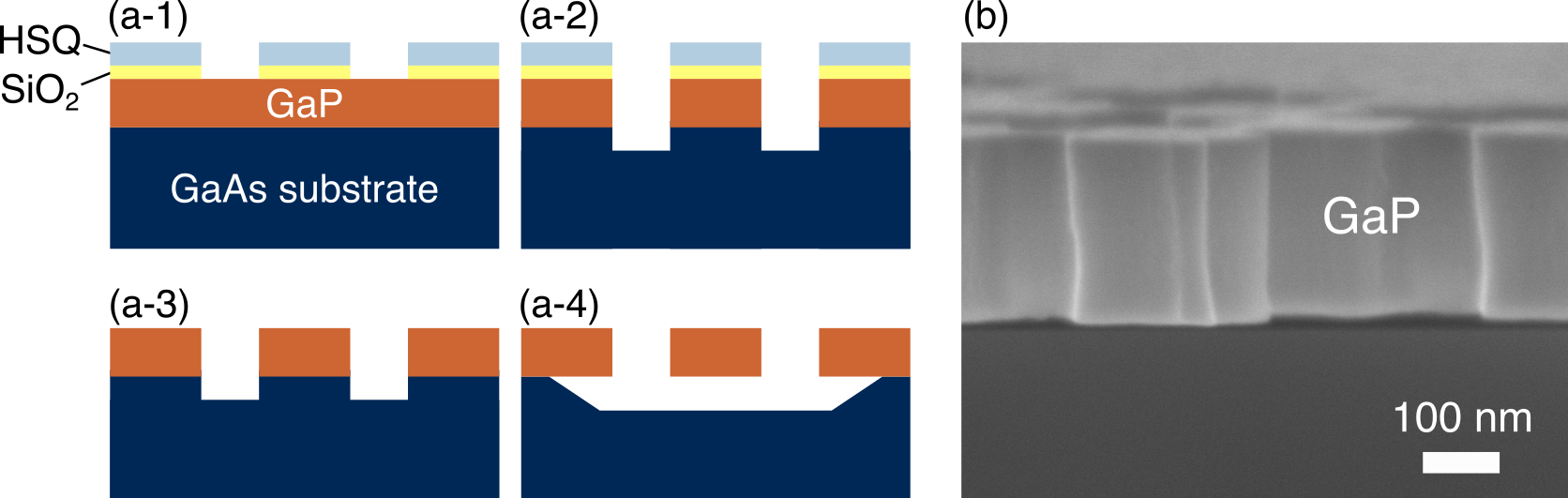}
\caption{
(a) Schematic of fabrication process. (a-1) Electron beam writer draws the OMC pattern on the HSQ resist which is adhered to GaP with a thin SiO$_2$ layer. (a-2) Dry etching with ICP plasma. (a-3) Removal of the resist. (a-4) Release of the device layer by chemically etching the substrate. (b) SEM image of a cleaved side wall of the device.
}
\label{fig:fab}
\end{figure}

Fig$.$ \ref{fig:fab}(a) draws schematic images of the fabrication process. Our chip comprises a 260 nm thick epitaxially grown GaP layer on top of a GaAs substrate. We deposit a thin (5 nm) SiO$_{2}$ layer to ensure the following HSQ resist firmly sticks to the GaP. The OMC patterns are drawn by a RAITH EBPG5200 electron beam (E-beam) writer and then dry etched with a SENTECH ICP-RIE SI500. The GaAs substrate is wet etched at the end by the chemical mixture of H$_{2}$SO$_{4}$ and H$_{2}$O$_{2}$ to suspend the device layer.
Fig$.$ \ref{fig:fab}(a) is an SEM image of a cleaved side view of the GaP device layer.
It shows that the sidewall of the fabricated device is not perfectly straight, possibly due to the E-beam writing or dry etching imperfection.

\begin{figure}[ht!]
\centering
\includegraphics[width=0.8\linewidth]{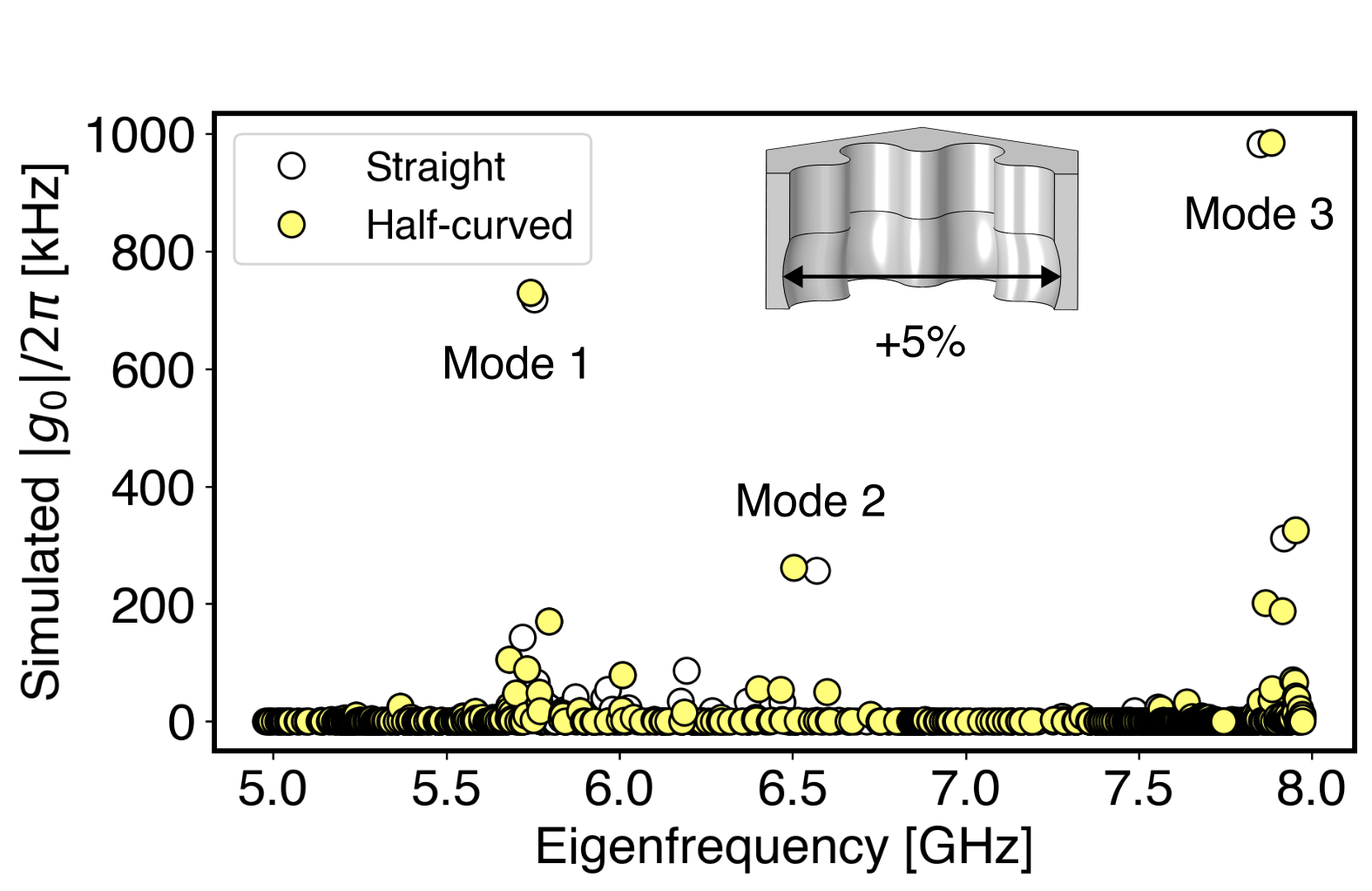}
\caption{
Simulated $g_{0}$ for the straight and curved sidewall snowflakes.
White (yellow) circles are the result of straight (hald-curved) side walls.
The inset is the cross section of the snowflake of the simulated model.
The curved part has maximum 5\% larger hole.
}
\label{fig:curved}
\end{figure}
We examine how much the fabrication imperfection affects the resulting optomechanical coupling rate.
We model the sidewall as shown in Fig$.$ \ref{fig:curved} which is made of straight wall for the upper half, and the quadratically curved upper side wall.
The widest part of the sidewall is 5\% wider than that of the straight part.
Fig$.$ \ref{fig:curved} shows the simulated vacuum coupling rate $g_0$ for the model.
Other geometry is identical to that of the main text.
We find that the negative effect from the curved side wall is quite limited for the 3 dominant modes as it almost only shifts the mechanical frequency slightly.
However it should be noted that the curve rate is not uniform over the entire device.
Snowflakes in one region have more deviation than that of other region, which might result in larger degradation of the coupling rate.

\section{Tapered fiber coupling}
\begin{figure}[ht!]
\centering
\includegraphics[width=0.8\linewidth]{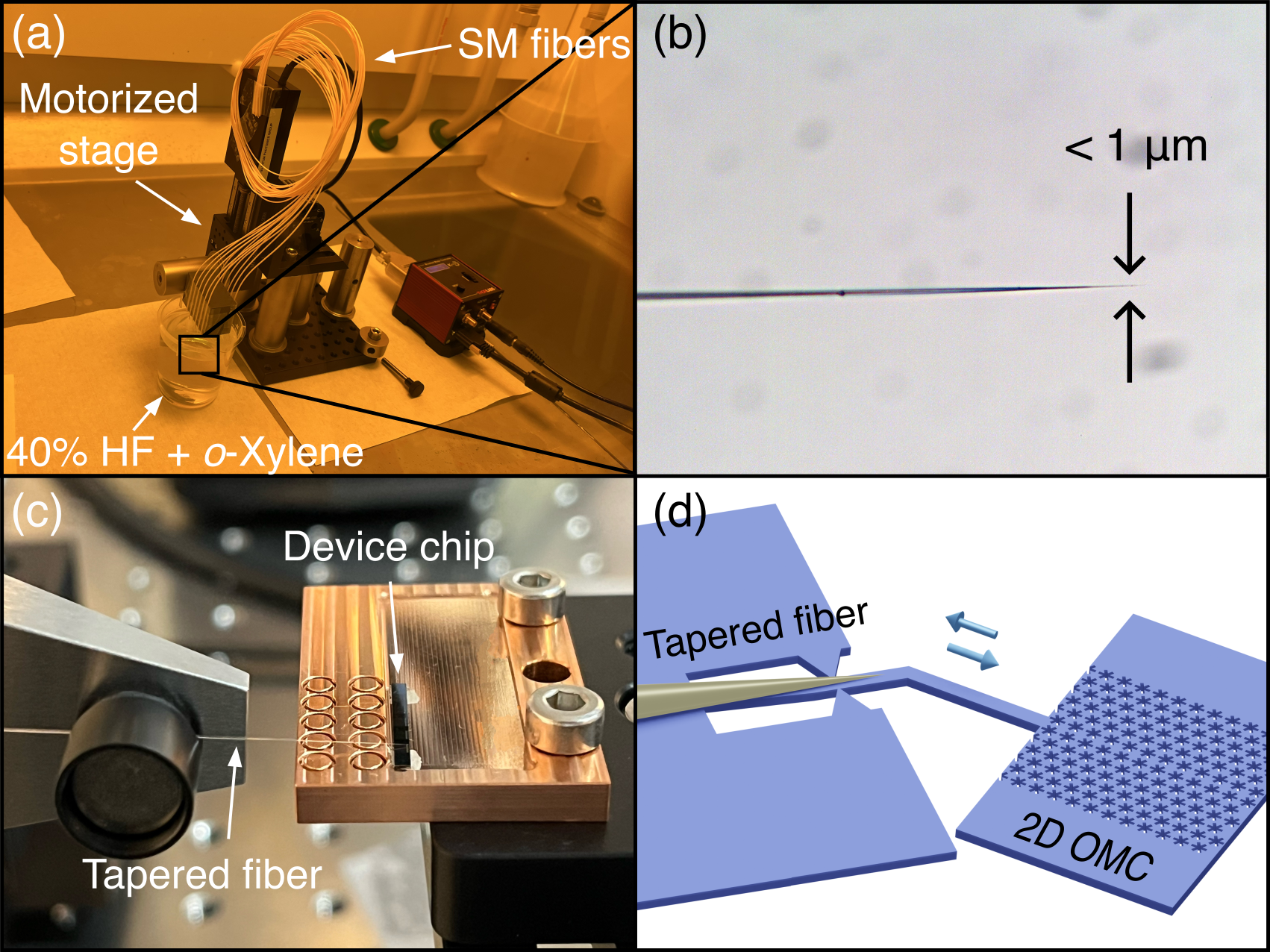}
\caption{Process of fiber preparation for the experiment. (a) Setup of fiber etching. Single-mode fibers are mobilized by a motorized stage which immerses them in 40\% HF+$o$-Xylene solution. (b) Microscope image of resulting fiber tip. (c) Picture of our fiber coupling setup. The device chip is mounted on a piezo positioning stage. (d) Schematic of the fiber-waveguide coupling configuration.}
\label{fig:fiber}
\end{figure}

We employ the evanescent fiber coupling approach\cite{tiecke_efficient_2015,burek_fiber-coupled_2017} with a chemically etched single-mode fiber tip for optical access to the device.
Fig$.$ \ref{fig:fiber}(a) is a photo of our setup for the fiber etching process. End-cut single-mode optical fibers are mounted to a motorized translational stage whose motion is digitally controlled.
We use 40$\%$ hydrofluoric (HF) acid solution covered by $o$-Xylene liquid preventing HF solution from evaporating.
The fiber tips are immersed in the solution by roughly 4 cm.
Then they are pulled up at a constant speed of $\sim$7 $\mu$m/s resulting in cone-shaped tips < 1 $\mu$m tip size shown in \ref{fig:fiber}(b).
The process is completed by cleaning the fiber tips by immersing them in an isopropanol solution for 2 mins.

During experiments, our test device is glued on a copper sample holder with silver conductive adhesive shown in Fig$.$ \ref{fig:fiber}(c).
We fix a tapered fiber on the manual translational stage with a v-groove to hold it. Then the sample is carefully positioned by piezo positioners.
We monitor the reflection intensity while adjusting the position and maximize the fiber-waveguide coupling.
Fig$.$ \ref{fig:fiber}(d) is a schematic image of the fiber-waveguide alignment.
The fiber tip touches the tapered waveguide which is bent 45 degrees so the fiber does not disturb our OMC.
By comparing the input and output light intensity, we estimate the coupling efficiency reaches > 60\%, which could be enhanced by optimizing the waveguide shape.
This configuration can reduce the vibrational noise by fibers, which can be significant at low temperatures and make the cryogenic experiments easier as no further alignment is needed.

\section{Polarization mismatch effect on vacuum coupling measurement}
\begin{figure}[ht!]
\centering
\includegraphics[width=0.8\linewidth]{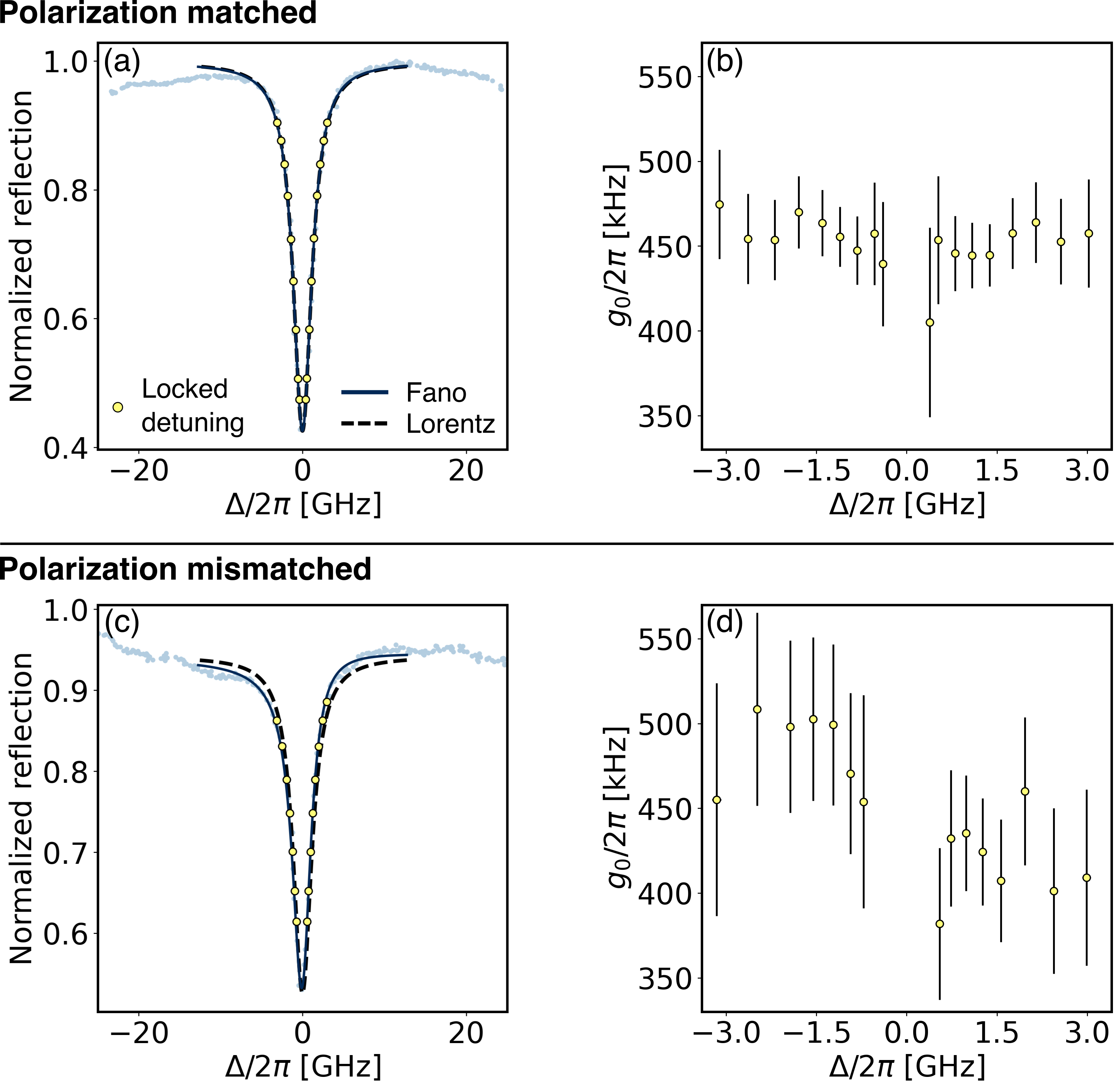}
\caption{Comparison of vacuum optomechanical coupling rate measurements under different polarization conditions. (a,b) Optical spectrum and resulting $g_{0}$ from matched polarization. The solid (dashed) curve is a Fano (Lorentz) fit. (c,d) Results from mismatched polarization.}
\label{fig:polarization}
\end{figure}
The calibration tone method to characterize $g_{0}$ can be easily degraded by the polarization mismatch of the input field from the waveguide mode.
This is due to the parasitic back reflection at the fiber-waveguide section which creates a sinusoidal optical spectrum background.
The back reflection causes interference between the light back-reflected and the one from the OMC, which changes the modulation depth of the calibration frequency component according to the shape of the optical spectrum.
To carry out the characterization in a low back-reflection effect, it is crucial to control the fiber position and the polarization precisely so the optical spectrum is as symmetric as possible.
For a polarization-matched condition, where the optical spectrum is symmetric Fig$.$ \ref{fig:polarization}(a), the calibration presents consistent $g_{0}$ over detuning Fig$.$ \ref{fig:polarization}(b).
Here, the yellow dots represent the locked laser detuning. However, under the polarization mismatched condition Fig$.$ \ref{fig:polarization}(c) where optical lineshape is not perfectly symmetric, estimated $g_0$ deviates from the intrinsic value as in Fig$.$ \ref{fig:polarization}(d), even without dynamical back-action.
This condition significantly depends on fiber position relative to the waveguide.

\section{Beam splitter model}
\begin{figure}[htbp]
\centering
\includegraphics[width=0.7\linewidth]{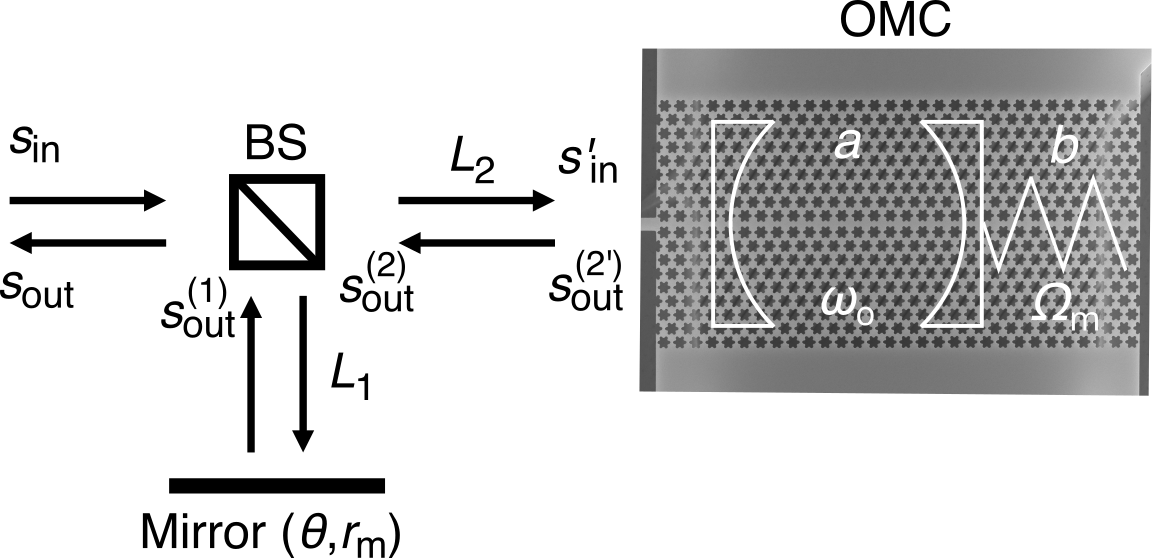}
\caption{Schematic image of the model. The beamsplitter (BS) has transmission (reflection) coefficient $t$ $(r)$. $L_{1}$ $(L_2)$ is the path length to mirror (OMC). The mirror has a reflection coefficient $re^{i\theta}$. The OMC possesses optical (mechanical) mode denoted by $a$ $(b)$ with the eigenfrequency $\omega_{\mathrm{o}}$ $(\Omega_{\mathrm{m}})$.}
\label{fig:beam-splitter-model}
\end{figure}
Here we consider the origin of the deviation that appears in the calibration tone method to determine $g_0$. We assume there is a residual back reflection from the fiber-waveguide coupling. We consider a model composed of a beam splitter (BS) and a mirror as well as our optomechanical crystal (OMC) that is drawn in Fig$.$ \ref{fig:beam-splitter-model}.

The phase-modulated input field is expressed as:
\begin{align}
\label{eq:input}
    s_{\mathrm{in}} = \left( s_{0} + s_{\mathrm{c}}e^{-i\Omega_{\mathrm{c}}t} - s_{\mathrm{c}}e^{+i\Omega_{\mathrm{c}}t}\right)e^{-i\omega_{\mathrm{L}}t},
\end{align}
where we define the laser frequency $\omega_{\mathrm{L}}$ and modulation frequency $\Omega_{\mathrm{c}}$.
Here we assume the amplitude of each frequency component to be real $s_{0},s_{\mathrm{c}}\in \mathbb{R}$.
The Langevin equation of motion for the optical annihilation operator $a$ is
\begin{align}
\label{eq:lang}
    \frac{d}{dt}a = -\left( i \omega_{\mathrm{o}} + \frac{\kappa}{2} \right)a + ig_{0}a\left( b^{\dagger} + b \right) + \sqrt{\kappa_{\mathrm{ex}}}s_{\mathrm{in}},
\end{align}
where $b$ is the annihilation operator of the mechanical mode.
Here, we take a Fourier-like expansion of optical and mechanical operators\cite{woolley_two-mode_2013,Woolley-2014},
\begin{align}
    &a = \left( a_{0} + a_{-}e^{-i\omega t} + a_{+}e^{i\omega t}  \right) e^{-i\omega_{\mathrm{L}} t},\\
    & b^{\dagger} + b= x = x_0 + x_{\mathrm{m}}e^{-i\Omega_{\mathrm{m}} t} + x_{\mathrm{m}}^{*}e^{i\Omega_{\mathrm{m}} t}.
\end{align}
where $x=b^{\dagger}+b$ is a unitless mechanical displacement. We explicitly assume that the mechanical motion has only a frequency component at $\Omega_{\mathrm{m}}$ by postulating there is no overlap between mechanical and calibration tone ($|\Omega_{\mathrm{m}}-\Omega_{\mathrm{c}}|\gg\Gamma_{\mathrm{m}}$). 
Inserting the above expression into the equation of motion and solving for the steady states, we get
\begin{align}
    &\bar{a}_0 = \sqrt{\kappa_{\mathrm{ex}}}s_0  \chi(0),  \\
    &\bar{a}_{-}(\omega)  = \begin{cases}
    \sqrt{\kappa_{\mathrm{ex}}}s_{\mathrm{c}} \chi(\Omega_{\mathrm{c}}) & \text{\ \  for $\ \ \ \ \omega=\Omega_{\mathrm{c}}$}\\
    i\sqrt{\kappa_{\mathrm{ex}}}g_{0}x_{\mathrm{m}}s_{0}\chi(0)\chi(\Omega_{\mathrm{m}}) & \text{\ \  for $\ \ \ \ \omega= \Omega_{\mathrm{m}}$ }
  \end{cases},\\
    &\bar{a}_{+}(\omega) = \begin{cases}
    -\sqrt{\kappa_{\mathrm{ex}}} s_{\mathrm{c}} \chi(-\Omega_{\mathrm{c}}) & \text{for $\ \ \ \ \omega=\Omega_{\mathrm{c}}$}\\
    i\sqrt{\kappa_{\mathrm{ex}}}g_{0}x_{\mathrm{m}}^{*}s_{0}\chi(0)\chi(-\Omega_{\mathrm{m}}) & \text{for $\ \ \ \ \omega = \Omega_{\mathrm{m}}$ }
    \end{cases},
\end{align}
where we define the optical susceptibility
\begin{align}
    \chi(\omega) \coloneq \frac{1}{\kappa/2 - i(\Delta + \omega)}.
\end{align}
Here we used the displaced detuning $\Delta' \coloneq \Delta + g_{0}x_{0}$ from the bare detuning $\Delta\coloneq \omega_{\mathrm{L}}-\omega_{\mathrm{o}}$ to account for the constant optical frequency shift, then redefine the detuning as $\Delta' \rightarrow \Delta$.
Now we introduce the two paths; one leads to a mirror which corresponds to residual back reflection from the imperfect fiber-waveguide coupling or waveguide itself, and the other to our OMC.
The output field from the mirror path $s_{\mathrm{out}}^{(1)}$ becomes
\begin{align}
    s_{\mathrm{out}}^{(1)} = r r_{\mathrm{m}}e^{i\theta}e^{2ikL_{1}}\left( s_{0} + s_{\mathrm{c}}e^{-i(\Omega_{\mathrm{c}}t-2\phi_{1}({\Omega_{\mathrm{c}}}))} - s_{\mathrm{c}}e^{i(\Omega_{\mathrm{c}}t-2\phi_{1}({\Omega_{\mathrm{c}}}))}  \right) e^{-i\omega_{\mathrm{L}} t},
\end{align}
where we define
\begin{align}
    \phi_{j}(\Omega) &\coloneq \frac{nL_{j}}{c}\Omega\ \ \ \ \ \mathrm{for}\ \ \ \ i=1,2,\\
    k &\coloneq \frac{n\omega_{\mathrm{L}}}{c}  
\end{align}
and refractive index $n$, phase shift by the mirror $\theta$, reflectance of the mirror $r_{\mathrm{m}}$, path length of each path $L_{j=1,2}$.

On the other hand, the fields impinging on the OMC obtain phase shifts in addition to those of eq.(\ref{eq:input}), by traveling the distance $L_2$. 
These phase shifts amount to:
\begin{align}
\label{eq:shifted_input0}
    s_0 e^{-i\omega_{\mathrm{L}} t} &\rightarrow s_{0} e^{-i\omega_{\mathrm{L}} t}e^{ikL_{2}},\\
\label{eq:shifted_input-}
    s_{\mathrm{c}} e^{-i(\omega_{\mathrm{L}}+\Omega_{\mathrm{c}}) t} &\rightarrow s_{\mathrm{c}} e^{-i(\omega_{\mathrm{L}}+\Omega_{\mathrm{c}}) t}e^{i(kL_{2}+\phi_{2}(\Omega_{\mathrm{c}}))},\\
\label{eq:shifted_input+}
    -s_{\mathrm{c}} e^{-i(\omega_{\mathrm{L}}-\Omega_{\mathrm{c}}) t} &\rightarrow -s_{\mathrm{c}} e^{-i(\omega_{\mathrm{L}}-\Omega_{\mathrm{c}}) t}e^{i(kL_{2}-\phi_{2}(\Omega_{\mathrm{c}}))}.
\end{align}
Therefore, the phase-shifted input field at the OMC, $s_{\mathrm{in}}^{'}$, is given by
\begin{align}
\label{eq:input-OMC}
    s_{\mathrm{in}}^{'} = te^{ikL_{2}}\left( s_{0} + s_{\mathrm{c}}e^{i\phi_{2}(\Omega_{\mathrm{c}})}e^{-i\Omega_{\mathrm{c}}t} - s_{\mathrm{c}}e^{-i\phi_{2}(\Omega_{\mathrm{c}})}e^{+i\Omega_{\mathrm{c}}t}\right)e^{-i\omega_{\mathrm{L}}t},
\end{align}
These modify the steady-state optical fields to:
\begin{align}
    &\bar{a}_0 \rightarrow t\sqrt{\kappa_{\mathrm{ex}}} \chi(0) s_{0} e^{ikL_{2}}, \\
    &\bar{a}_{-}(\omega)  \rightarrow \begin{cases}
    t\sqrt{\kappa_{\mathrm{ex}}}  s_{\mathrm{c}}\chi(\Omega_{\mathrm{c}})e^{i(kL_2+\phi_{2}(\Omega_{\mathrm{c}}))} &\  \text{      for $\ \ \ \ \omega=\Omega_{\mathrm{c}}$}\\
    it\sqrt{\kappa_{\mathrm{ex}}}g_{0}x_{\mathrm{m}}s_{0}\chi(0)\chi(\Omega_{\mathrm{m}})e^{ikL_2} &\  \text{       for $\ \ \ \ \omega = \Omega_{\mathrm{m}}$ }
  \end{cases},\\
    &\bar{a}_{+}(\omega)  \rightarrow \begin{cases}
    -t\sqrt{\kappa_{\mathrm{ex}}}s_{\mathrm{c}}\chi(-\Omega_{\mathrm{c}})e^{i(kL_2-\phi_{2}(\Omega_{\mathrm{c}}))} & \text{for $\ \ \ \ \omega=\Omega_{\mathrm{c}}$}\\    it\sqrt{\kappa_{\mathrm{ex}}}g_{0}x_{\mathrm{m}}^{*}s_{0}\chi(0)\chi(-\Omega_{\mathrm{m}})e^{ikL_2} & \text{for $\ \ \ \ \omega = \Omega_{\mathrm{m}}$ }
    \end{cases}.
\end{align}
From the input-output theory and taking the transmission of the beamsplitter into account, the output field right after the OMC $s_{\mathrm{out}}^{(2')}$ is,
\begin{align}
    s_{\mathrm{out}}^{(2')}(\omega) = s'_{\mathrm{in}} - \sqrt{\kappa_{\mathrm{ex}}}a(\omega),
\end{align}
where $t=\sqrt{1-r^2}$.
Therefore, we can calculate the calibration tone component of the output field from the OMC just before the beam splitter,
\begin{align}
\begin{split}
   s_{\mathrm{out}}^{(2)}(\Omega_{\mathrm{c}}) &= t s_{\mathrm{c}}e^{i(-\omega_{\mathrm{L}}t+2kL_2)}\Big[ \big\{ 
1-\kappa_{\mathrm{ex}}\chi(\Omega_{\mathrm{c}}) \big\}e^{-i(\Omega_{\mathrm{c}}t-2\phi_{2}(\Omega_{\mathrm{c}}))}  \\
&\qquad\qquad\qquad\qquad - \big\{ 
1-\kappa_{\mathrm{ex}}\chi(-\Omega_{\mathrm{c}}) \big\}e^{i(\Omega_{\mathrm{c}}t-2\phi_{2}(\Omega_{\mathrm{c}}))} \Big],
\end{split}
\end{align}
and the mechanical component,
\begin{align}
    \begin{split}
         s_{\mathrm{out}}^{(2)}(\Omega_{\mathrm{m}}) &= - it\kappa_{\mathrm{ex}}g_{0}\chi(0)s_0 e^{i(-\omega_{\mathrm{L}}t+2kL_2)} \Big[  
x_{\mathrm{m}}\chi(\Omega_{\mathrm{m}})e^{-i(\Omega_{\mathrm{m}}t-\phi_{2}(\Omega_{\mathrm{m}}))}\\
&\qquad\qquad\qquad\qquad\qquad\qquad\quad + x_{\mathrm{m}}^{*}\chi(-\Omega_{\mathrm{m}})e^{i(\Omega_{\mathrm{m}}t-\phi_{2}(\Omega_{\mathrm{m}}))} \Big],
    \end{split}
\end{align}
and finally the carrier frequency component,
\begin{align}
    s_{\mathrm{out}}^{(2)}(0) = t e^{i(-\omega_{\mathrm{L}}t+2kL_2)}\big\{ 
1-\kappa_{\mathrm{ex}}\chi(0) \big\}s_0.
\end{align}
The total output field we measure is the combined field after the beam splitter which is,
\begin{align}
    s_{\mathrm{out}} = rs_{\mathrm{out}}^{(1)}+ts_{\mathrm{out}}^{(2)}.
\end{align}
From the above expressions for $s_{\mathrm{out}}^{(1)}$ and $s_{\mathrm{out}}^{(2)}$, we have the total output field for each frequency component as,
\begin{align}
    e^{i(\omega_{\mathrm{L}}t-2kL_2)}s_{\mathrm{out}}(0) &= \left[ r^2 r_{\mathrm{m}}e^{i(\theta+2k\Delta L)} + t^2 \big\{1-\kappa_{\mathrm{ex}}\chi(0)\big\} \right]s_0 \eqcolon A,\\
    \begin{split}
    e^{i(\omega_{\mathrm{L}}t-2kL_2)}s_{\mathrm{out}}(\Omega_{\mathrm{c}}) &= \left[ r^2 r_{\mathrm{m}}e^{i(\theta+2k\Delta L + 2\phi_{1}(\Omega_{\mathrm{c}}))} + t^2 \big\{1-\kappa_{\mathrm{ex}}\chi(\Omega_{\mathrm{c}})\big\}e^{2i\phi_{2}(\Omega_{\mathrm{c}})} \right]s_{\mathrm{c}}e^{-i\Omega_{\mathrm{c}}t}\\
        &\ \ \ \ - \left[ r^2 r_{\mathrm{m}}e^{i(\theta+2k\Delta L - 2\phi_{1}(\Omega_{\mathrm{c}}))} + t^2 \big\{1-\kappa_{\mathrm{ex}}\chi(-\Omega_{\mathrm{c}})\big\}e^{-2i\phi_{2}(\Omega_{\mathrm{c}})} \right]s_{\mathrm{c}}e^{i\Omega_{\mathrm{c}}t}\\
        & \eqcolon B_{\mathrm{c}}e^{-i\Omega_{\mathrm{c}}t} + C_{\mathrm{c}}e^{i\Omega_{\mathrm{c}}t},
    \end{split}\\
    \begin{split}
        e^{i(\omega_{\mathrm{L}}t-2kL_2)}s_{\mathrm{out}}(\Omega_{\mathrm{m}})&=- i t^2 \kappa_{\mathrm{ex}}g_0 \chi(0)s_{\mathrm{c}}\Big\{ x_{\mathrm{m}}\chi(\Omega_{\mathrm{m}})e^{-i(\Omega_{\mathrm{m}}t-\phi_{2}((\Omega_{\mathrm{m}}))} \\
        & \qquad\qquad\qquad\qquad  +x_{\mathrm{m}}^{*}\chi(-\Omega_{\mathrm{m}})e^{i(\Omega_{\mathrm{m}}t-\phi_{2}((\Omega_{\mathrm{m}}))} \Big\}
    \end{split}\\
    & \eqcolon B_{\mathrm{m}}e^{-i\Omega_{\mathrm{m}}t} + C_{\mathrm{m}}e^{i\Omega_{\mathrm{m}}t}.
\end{align}
where we define $\Delta L \coloneq L_1 - L_2$.
The calibration frequency $\Omega_{\mathrm{c}}$ component of intensity modulation of the output field is
\begin{align}
    \left| s_{\mathrm{out}}\right|^2(\Omega_{\mathrm{c}})  &=s_{\mathrm{out}}(0)s_{\mathrm{out}}^{*} (\Omega_{\mathrm{c}}) + \text{h.c.}\\
    &= \left(A^{*}B_{\mathrm{c}}+AC_{\mathrm{c}}^{*}\right) e^{-i\Omega_{\mathrm{c}}t} + \text{h.c.}
\end{align}
Applying the same procedure for the mechanical frequency component, we have
\begin{align}
    \left| s_{\mathrm{out}}\right|^2(\Omega_{\mathrm{m}})  &=s_{\mathrm{out}}(0)s_{\mathrm{out}}^{*} (\Omega_{\mathrm{m}}) + \text{h.c.}\\
    &= \left(A^{*}B_{\mathrm{m}}+AC_{\mathrm{m}}^{*}\right) e^{-i\Omega_{\mathrm{m}}t} + \text{h.c.}
\end{align}
To estimate the impact on $g_0$, we define the mechanics/calibration power ratio $\eta_{g}(\Delta)$:
\begin{align}
\label{eq:ratio}
    \eta_{g}(\Delta) \coloneq \left| \frac{A^{*}B_{\mathrm{m}}+AC_{\mathrm{m}}^{*}}{A^{*}B_{\mathrm{c}}+AC_{\mathrm{c}}^{*}} \right|^2,
\end{align}
which is proportional to the measured $g_0$.
We assume that the dynamical backaction is small enough so the mechanical linewidth and mechanical peak hight are unchanged over detuning.

\begin{figure}[ht!]
\centering
\includegraphics[width=0.9\linewidth]{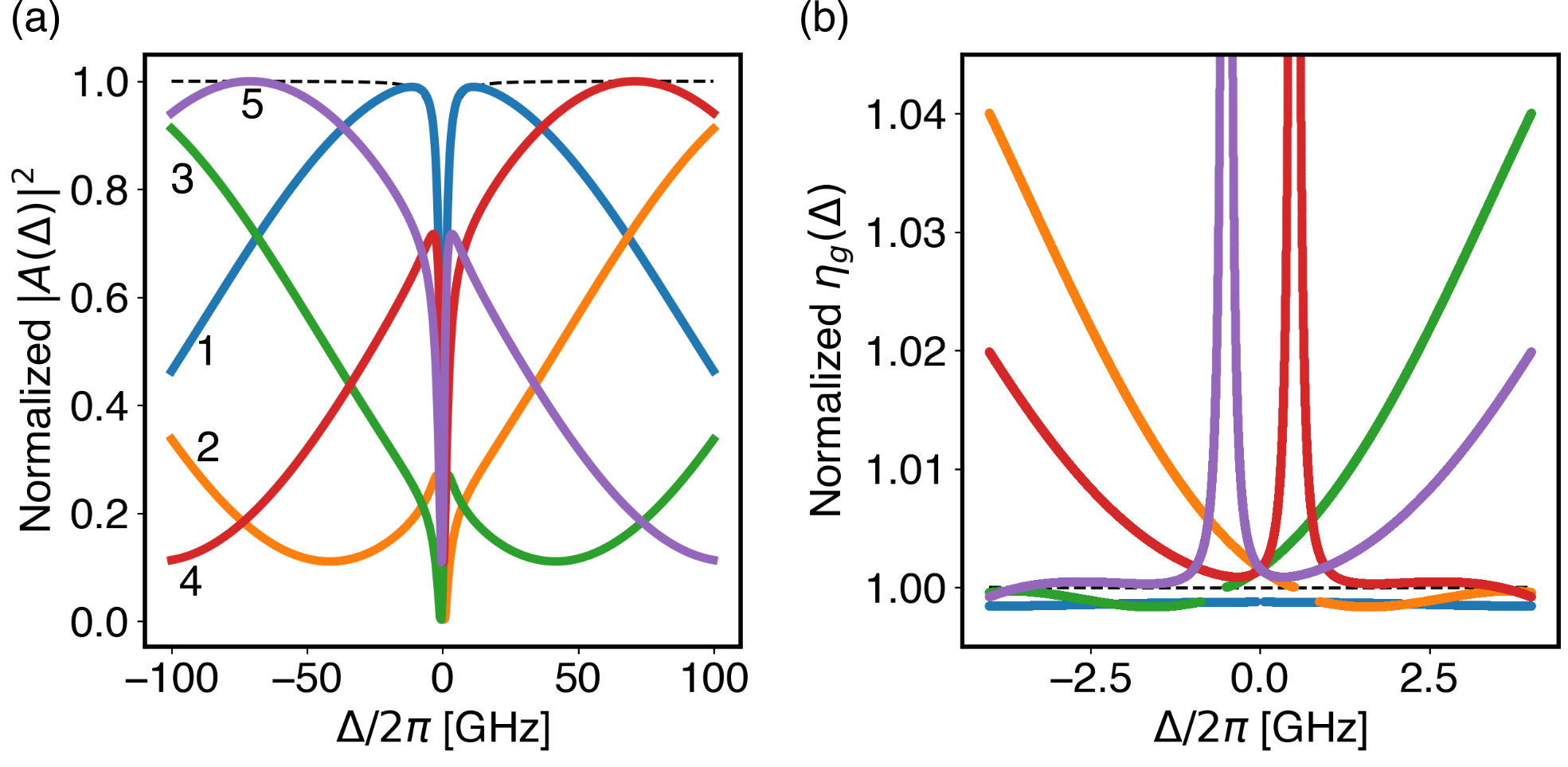}
\caption{(a)Reflection coefficient and (b) Mechanical/calibration ratio $\eta_{g}$ over detuning. Used parameters are $\omega_{\mathrm{o}}/2\pi$ = 195.55 THz, $n$ = 3.05, $\Delta L$ = -140 $\mu$m, $\kappa_0/2\pi$ = 1.5 GHz, $\kappa_{\mathrm{ex}}/2\pi$ = 1 GHz, $\Omega_{\mathrm{m}}/2\pi$ = 7.65 GHz, and $\Omega_{\mathrm{c}}/2\pi$ = 7.65 GHz. 
For curves 1 to 5, $\theta-2n\omega_{\mathrm{L}}\Delta L/c$ = $0,0.77\pi,-0.77\pi,0.4\pi,-0.4\pi$
respectively. Dashed black curves are references with $r=0$.}
\label{fig:theory_plot}
\end{figure}
Fig$.$ \ref{fig:theory_plot} shows plots of $|A(\Delta)|^2$ and $\eta_{g}(\Delta)$ with parameters of $\omega_{\mathrm{o}}/2\pi$ = 195.55 THz, $n$ = 3.05, $\Delta L$ = -140 $\mu$m, $\kappa_0/2\pi$ = 1.5 GHz, $\kappa_{\mathrm{ex}}/2\pi$ = 1 GHz, $\Omega_{\mathrm{m}}/2\pi$ = 7.65 GHz, and $\Omega_{\mathrm{c}}/2\pi$ = 7.65 GHz.
Fig$.$ \ref{fig:theory_plot}(a) corresponds to the optical spectra in polarization-matched (unmatched) conditions for curve 1 (curve 2-5).
(b) plots the $\eta_{g}(\Delta)$ for each condition which is proportional to the measured $g_{0}$\cite{Gorodetsky2010}.
As it shows, the measured $g_{0}$ depends on laser detuning if the optical spectrum is in polarization unmatched condition regardless of dynamical backaction. Here we skip the points where the denominator of eq.(\ref{eq:ratio}) is small corresponding to the calibration tone being too small to detect.
With the realistic parameters of our OMC device, we find the measured $g_0$ can change roughly 4\% over $\pm$ 4 GHz of detuning depending hugely on the phase $\theta$.
This deviation is still smaller than that we obtain from the measurements.
A possible reason is that our model is oversimplified and other mechanisms give larger phase differences in the mechanical and calibration components.
In the actual device, we may speculate the reflectance $r$ of the BS is frequency-dependent or that there are multiple BSs, which may show more complex behavior.
Still, this result strongly suggests that the parasitic back reflection in the waveguide affects the measurement on $g_0$.
Further optimization of the fiber-waveguide coupling will suppress the back reflection of propagating light.

In addition to the $g_0$ deviation, this model predicts the Fano-like reflection spectrum observed in Fig.~\ref{fig:polarization}(c).
To this end, we write $|s_\mathrm{out}|^2$, disregarding the calibration tone and mechanical sidebands, i.e. retaining only 
\begin{align}\label{eq:A^2}
    |A|^2 =&\ |s_0|^2 |r^2 r_m e^{i(\theta + 2k\Delta L)} + t^2\left(1- \kappa_\mathrm{ex} \chi(0) \right)|^2 \\
    =&\  |s_0|^2 \bigg(
    \left( |r^2r_m|+|t|^2\cos(\theta + 2 k \Delta L) \right)^2 \\
    &- 2|t|^4 \kappa(\eta_c-\eta_c^2)\frac{\frac{\kappa}{2} \left(1  +\frac{r^2r_m}{|t|^2(1-\eta_c)}\cos(\theta + 2k\Delta L) \right) +\Delta \frac{r^2r_m}{|t|^2(1-\eta_c)} \sin(\theta + 2k \Delta L) }{\left(\frac{\kappa}{2}\right)^2 + \Delta^2}
    \bigg).
\end{align}
Now, eq.~(1) in the main text follows from eq.~(\ref{eq:A^2}) by identifying
\begin{align}
    h&=|s_0|^2 \left(|r^2r_m|+|t|^2\cos(\theta + 2 k \Delta L) \right)^2,\\
    A&=2|s_0|^2|t|^4  \kappa(\eta_c-\eta_c^2),\\
    q &= \frac{r^2r_m}{|t|^2(1-\eta_c)} \sin(\theta+2k\Delta L)
\end{align}
by requiring $ \cos(\theta+2k\Delta L) = -\sin(\theta+2k\Delta L)^2$.
As is shown in Fig.~\ref{fig:theory_plot}(a), eq.~(\ref{eq:A^2}) exhibit similar Fano-like reflection spectra for a range of interferometer phases $\theta.$
In practice, we further approximate $h\approx \mathrm{const.}$ by restricting our fitting region to the vicinity of the cavity resonance.



\bibliography{Room-Temp_GaP}

\begin{thebibliography}{10}
\newcommand{\enquote}[1]{``#1''}

\bibitem{gisler2024enhancingmembranebasedscanningforce}
T.~Gisler, D.~Hälg, V.~Dumont, S.~Misra, L.~Catalini, E.~C. Langman,
  A.~Schliesser, C.~L. Degen, and A.~Eichler, \enquote{Enhancing membrane-based
  scanning force microscopy through an optical cavity,}
  {\protect\JournalTitle{arXiv:2406.07171}}  (2024).

\bibitem{meesala_non-classical_2024}
S.~Meesala, S.~Wood, D.~Lake, P.~Chiappina, C.~Zhong, A.~D. Beyer, M.~D. Shaw,
  L.~Jiang, and O.~Painter, \enquote{Non-classical microwave–optical photon
  pair generation with a chip-scale transducer,} {\protect\JournalTitle{Nat.
  Phys.}} \textbf{20}, 871--877 (2024).

\bibitem{PhysRevLett.132.100802}
M.~B. Kristensen, N.~Kralj, E.~C. Langman, and A.~Schliesser,
  \enquote{Long-lived and efficient optomechanical memory for light,}
  {\protect\JournalTitle{Phys. Rev. Lett.}} \textbf{132}, 100802 (2024).

\bibitem{Andreas2020}
A.~Wallucks, I.~Marinkovi{\'c}, B.~Hensen, R.~Stockill, and S.~Gr{\"o}blacher,
  \enquote{A quantum memory at telecom wavelengths,}
  {\protect\JournalTitle{Nature Physics}} \textbf{16}, 772--777 (2020).

\bibitem{Fiaschi2021}
N.~Fiaschi, B.~Hensen, A.~Wallucks, R.~Benevides, J.~Li, T.~P. Alegre, and
  S.~Gr{\"o}blacher, \enquote{Optomechanical quantum teleportation,}
  {\protect\JournalTitle{Nature Photonics}} \textbf{15}, 817--821 (2021).

\bibitem{Briegel-repeater-1998}
H.-J. Briegel, W.~D\"ur, J.~I. Cirac, and P.~Zoller, \enquote{Quantum
  repeaters: The role of imperfect local operations in quantum communication,}
  {\protect\JournalTitle{Phys. Rev. Lett.}} \textbf{81}, 5932--5935 (1998).

\bibitem{pittaluga_600-km_2021}
M.~Pittaluga, M.~Minder, M.~Lucamarini, M.~Sanzaro, R.~I. Woodward, M.-J. Li,
  Z.~Yuan, and A.~J. Shields, \enquote{600-km repeater-like quantum
  communications with dual-band stabilization,} {\protect\JournalTitle{Nature
  Photonics}} \textbf{15}, 530--535 (2021).

\bibitem{julsgaard_experimental_2004}
B.~Julsgaard, J.~Sherson, J.~I. Cirac, J.~Flurášek, and E.~S. Polzik,
  \enquote{Experimental demonstration of quantum memory for light,}
  {\protect\JournalTitle{Nature}} \textbf{432}, 482--486 (2004).

\bibitem{yuan_experimental_2008}
Z.~S. Yuan, Y.~A. Chen, B.~Zhao, S.~Chen, J.~Schmiedmayer, and J.~W. Pan,
  \enquote{Experimental demonstration of a {BDCZ} quantum repeater node,}
  {\protect\JournalTitle{Nature}} \textbf{454}, 1098--1101 (2008).

\bibitem{pedersen_near_2020}
F.~T. Pedersen, Y.~Wang, C.~T. Olesen, S.~Scholz, A.~D. Wieck, A.~Ludwig, M.~C.
  Löbl, R.~J. Warburton, L.~Midolo, R.~Uppu, and P.~Lodahl, \enquote{Near
  transform-limited quantum dot linewidths in a broadband photonic crystal
  waveguide,} {\protect\JournalTitle{{ACS} Photonics}} \textbf{7}, 2343--2349
  (2020).

\bibitem{aspelmeyer_cavity_2014}
M.~Aspelmeyer, T.~J. Kippenberg, and F.~Marquardt, \enquote{Cavity
  optomechanics,} {\protect\JournalTitle{Reviews of Modern Physics}}
  \textbf{86}, 1391--1452 (2014).

\bibitem{jiang_optically_2023}
W.~Jiang, F.~M. Mayor, S.~Malik, R.~Van~Laer, T.~P. McKenna, R.~N. Patel, J.~D.
  Witmer, and A.~H. Safavi-Naeini, \enquote{Optically heralded microwave photon
  addition,} {\protect\JournalTitle{Nat. Phys.}} \textbf{19}, 1423--1428
  (2023).

\bibitem{HRen2020}
H.~Ren, M.~H. Matheny, G.~S. MacCabe, J.~Luo, H.~Pfeifer, M.~Mirhosseini, and
  O.~Painter, \enquote{Two-dimensional optomechanical crystal cavity with high
  quantum cooperativity,} {\protect\JournalTitle{Nature Communications}}
  \textbf{11}, 3373 (2020).

\bibitem{sonar_high-efficiency_2024}
S.~Sonar, U.~Hatipoglu, S.~Meesala, D.~Lake, H.~Ren, and O.~Painter,
  \enquote{High-efficiency low-noise optomechanical crystal photon-phonon
  transducers,} {\protect\JournalTitle{arXiv:2406.15701}}  (2024).

\bibitem{Schneider2019}
K.~Schneider, Y.~Baumgartner, S.~H{\"o}nl, P.~Welter, H.~Hahn, D.~J. Wilson,
  L.~Czornomaz, and P.~Seidler, \enquote{Optomechanics with one-dimensional
  gallium phosphide photonic crystal cavities,} {\protect\JournalTitle{Optica}}
  \textbf{6}, 577 (2019).

\bibitem{Stockill2019}
R.~Stockill, M.~Forsch, G.~Beaudoin, K.~Pantzas, I.~Sagnes, R.~Braive, and
  S.~Gr{\"o}blacher, \enquote{Gallium phosphide as a piezoelectric platform for
  quantum optomechanics,} {\protect\JournalTitle{Physical Review Letters}}
  \textbf{123} (2019).

\bibitem{Gorodetsky2010}
M.~L. Gorodetsky, A.~Schliesser, G.~Anetsberger, S.~Deleglise, and T.~J.
  Kippenberg, \enquote{Determination of the vacuum optomechanical coupling rate
  using frequency noise calibration,} {\protect\JournalTitle{Optics Express}}
  \textbf{18}, 23236 (2010).

\bibitem{Safavi-Naeini2014}
A.~H. Safavi-Naeini, J.~T. Hill, S.~Meenehan, J.~Chan, S.~Gr{\"o}blacher, and
  O.~Painter, \enquote{Two-dimensional phononic-photonic band gap
  optomechanical crystal cavity,} {\protect\JournalTitle{Physical Review
  Letters}} \textbf{112}, 1--5 (2014).

\bibitem{PhysRevApplied.21.014015}
R.~G. Povey, M.-H. Chou, G.~Andersson, C.~R. Conner, J.~Grebel, Y.~J. Joshi,
  J.~M. Miller, H.~Qiao, X.~Wu, H.~Yan, and A.~N. Cleland,
  \enquote{Two-dimensional optomechanical crystal resonator in gallium
  arsenide,} {\protect\JournalTitle{Phys. Rev. Appl.}} \textbf{21}, 014015
  (2024).

\bibitem{akahane_high-q_2003}
Y.~Akahane, T.~Asano, B.-S. Song, and S.~Noda, \enquote{High-{Q} photonic
  nanocavity in a two-dimensional photonic crystal,}
  {\protect\JournalTitle{Nature}} \textbf{425}, 944--947 (2003).

\bibitem{comsol}
COMSOL Multiphysics® v. 5.6. www.comsol.com. COMSOL AB, Stockholm, Sweden.

\bibitem{chan_optimized_2012}
J.~Chan, A.~H. Safavi-Naeini, J.~T. Hill, S.~Meenehan, and O.~Painter,
  \enquote{Optimized optomechanical crystal cavity with acoustic radiation
  shield,} {\protect\JournalTitle{Applied Physics Letters}} \textbf{101},
  081115 (2012).

\bibitem{balram_moving_2014}
K.~C. Balram, M.~Davan\c{c}o, J.~Y. Lim, J.~D. Song, and K.~Srinivasan,
  \enquote{Moving boundary and photoelastic coupling in gaas optomechanical
  resonators,} {\protect\JournalTitle{Optica}} \textbf{1}, 414--420 (2014).

\bibitem{mytsyk_elasto-optic_2015}
B.~G. Mytsyk, N.~M. Demyanyshyn, and O.~M. Sakharuk, \enquote{Elasto-optic
  effect anisotropy in gallium phosphide crystals,}
  {\protect\JournalTitle{Appl. Opt.}} \textbf{54}, 8546 (2015).

\bibitem{biegelsen_photoelastic_1974}
D.~K. Biegelsen, \enquote{Photoelastic {Tensor} of {Silicon} and the {Volume}
  {Dependence} of the {Average} {Gap},} {\protect\JournalTitle{Phys. Rev.
  Lett.}} \textbf{32}, 1196--1199 (1974).

\bibitem{tiecke_efficient_2015}
T.~G. Tiecke, K.~P. Nayak, J.~D. Thompson, T.~Peyronel, N.~P. de~Leon,
  V.~Vuletić, and M.~D. Lukin, \enquote{Efficient fiber-optical interface for
  nanophotonic devices,} {\protect\JournalTitle{Optica}} \textbf{2}, 70 (2015).

\bibitem{burek_fiber-coupled_2017}
M.~J. Burek, C.~Meuwly, R.~E. Evans, M.~K. Bhaskar, A.~Sipahigil, S.~Meesala,
  B.~MacHielse, D.~D. Sukachev, C.~T. Nguyen, J.~L. Pacheco, E.~Bielejec, M.~D.
  Lukin, and M.~Lončar, \enquote{Fiber-coupled diamond quantum nanophotonic
  interface,} {\protect\JournalTitle{Physical Review Applied}} \textbf{8},
  024026 (2017).

\bibitem{zeng_cryogenic_2023}
B.~Zeng, C.~De-Eknamkul, D.~Assumpcao, D.~Renaud, Z.~Wang, D.~Riedel, J.~Ha,
  C.~Robens, D.~Levonian, M.~Lukin, R.~Riedinger, M.~Bhaskar, D.~Sukachev,
  M.~Loncar, and B.~Machielse, \enquote{Cryogenic packaging of nanophotonic
  devices with a low coupling loss <1 {dB},} {\protect\JournalTitle{Applied
  Physics Letters}} \textbf{123}, 161106 (2023).

\bibitem{carmon_dynamical_2004}
T.~Carmon, L.~Yang, and K.~J. Vahala, \enquote{Dynamical thermal behavior and
  thermal self-stability of microcavities,} {\protect\JournalTitle{Optics
  Express}} \textbf{12}, 4742 (2004).

\bibitem{honl_microwave--optical_2022}
S.~Hönl, Y.~Popoff, D.~Caimi, A.~Beccari, T.~J. Kippenberg, and P.~Seidler,
  \enquote{Microwave-to-optical conversion with a gallium phosphide photonic
  crystal cavity,} {\protect\JournalTitle{Nat Commun}} \textbf{13}, 2065
  (2022).

\bibitem{forsch_microwave--optics_2020}
M.~Forsch, R.~Stockill, A.~Wallucks, I.~Marinković, C.~Gärtner, R.~A. Norte,
  F.~van Otten, A.~Fiore, K.~Srinivasan, and S.~Gröblacher,
  \enquote{Microwave-to-optics conversion using a mechanical oscillator in its
  quantum ground state,} {\protect\JournalTitle{Nature Physics}} \textbf{16},
  69--74 (2020).

\bibitem{woolley_two-mode_2013}
M.~J. Woolley and A.~A. Clerk, \enquote{Two-mode back-action-evading
  measurements in cavity optomechanics,} {\protect\JournalTitle{Physical Review
  A}} \textbf{87}, 1--24 (2013).

\bibitem{Woolley-2014}
M.~J. Woolley and A.~A. Clerk, \enquote{Two-mode squeezed states in cavity
  optomechanics via engineering of a single reservoir,}
  {\protect\JournalTitle{Phys. Rev. A}} \textbf{89}, 063805 (2014).

\end{thebibliography}
\end{document}